\documentclass[twocolumn]{aastex631}
\usepackage{hyperref}
\usepackage{comment}

\begin{document}

\title{Next-Generation Atmosphere Models for Giant Planets with Application to Coupled Interior Composition and Spectral Evolution I: Cloudless Models with Equilibrium Chemistry}

\shorttitle{Atmosphere Models for Giant Planets}
\shortauthors{Chen et al.}


\author[0000-0003-3792-2888]{Yi-Xian Chen}
\affiliation{Department of Astrophysical Sciences, Princeton University,  4 Ivy Lane, Princeton, NJ 08544, USA}

\author[0000-0001-6708-3427]{Roberto Tejada Arevalo}
\affiliation{Department of Astrophysical Sciences, Princeton University,  4 Ivy Lane, Princeton, NJ 08544, USA}

\author[0000-0002-3099-5024]{Adam Burrows}
\affiliation{Department of Astrophysical Sciences, Princeton University,  4 Ivy Lane, Princeton, NJ 08544, USA}

\author[0000-0001-6635-5080]{Ankan Sur}
\affiliation{Department of Astrophysical Sciences, Princeton University,  4 Ivy Lane, Princeton, NJ 08544, USA}






\begin{abstract}
We present updated atmosphere models designed for calculating the post-formation evolution and cooling of giant planets with masses between $0.3$ and $10$ $M_J$. 
Our tables provide the entropy in the convective region at the base of the atmosphere, temperature ($T$)–pressure ($P$) profiles, and emergent spectra for atmospheres calculated using the radiative transfer code \texttt{CoolTLusty} for $T_{\mathrm{eff}}$s over the range 100 to 1400 Kelvin and log$_{10}$($g$) from 2.8 to 4.4 ($cgs$) with the latest opacities and equations of state. Each spectrum and thermal profile is calculated using line-by-line opacity sampling. We construct tables at 3 different metallicities ($Z = 1, 3.16, 10 Z_\odot$) and 2 different helium fractions ($Y=0.15, 0.275$), with the improvement that we adopt a metal-inclusive EOS that treats heavy elements consistently with the opacity metallicity (rather than folding it into an effective $Y$). The result is tables that accommodate both changes in $Y$ due to helium rain and potential variations in $Z$ during envelope evolution. 
We present a comparison between T-P profiles, modeled spectra, and evolutionary tracks, and find that on-the-fly interpolation of boundary conditions in atmospheric composition 
has a notable impact on the late stages of giant planet evolution, 
altering the timing of helium rain and therefore the subsequent cooling history and atmospheric helium depletion.
We also provide a ready-to-use toolkit that generates spectra and boundary conditions via efficient interpolation across the four-dimensional parameter space $(T_{\rm eff}, \log_{10} g, Y, Z)$, which is useful for post-processing evolutionary tracks to produce fully time-resolved spectral evolutions.
\end{abstract}



\section{Introduction} \label{sec:intro}

The cooling and contraction of the 
interiors of giant planets following their formation 
are controlled by energy loss from their comparatively thin outer atmospheres.
Accurate and efficient long-term evolutionary calculations, 
therefore, require a grid of models that self-consistently determine the atmospheric temperature–pressure (T-P) structures for a range of effective temperatures, 
surface gravities, 
and atmospheric compositions. 
Such atmospheric grids set the outer boundary conditions for evolutionary models, 
{enabling the planet’s thermodynamic state and composition to be tracked over time.}

Early studies of the thermal evolution of Jupiter and Saturn (e.g., \citealt{Graboske1975,Pollack1977,Hubbard1977}) relied upon boundary conditions derived from highly simplified radiative-transfer treatments \citep{Pollack1973}.
Following the discovery of hot Jupiters \citep{Mayor1995}, 
extensive new atmosphere grids were constructed to explore a broader parameter space relevant to extrasolar gas giants and brown dwarfs \citep{Burrows1997,Burrows2001}. 
{Subsequent developments incorporated improved descriptions of stellar irradiation \citep{Sudarsky2000,Burrows2003,Burrows2006,Fortney2011}, updated opacity databases \citep{SharpBurrows2007,Lacy2023}, and more realistic cloud physics \citep{Ackerman2001,Burrows2006cloud,Marley2010,Chen2023,Morley2024,Huang2024,mang2026picaso}.} 

For systematic coverage of isolated sub-stellar atmospheres, the leading frameworks are currently the Sonora–Bobcat grid \citep{Marley2021,Morley2024} and the ATMO2020 models \citep{Phillips2020}.
The growing sample of directly imaged and characterized planetary companions \citep{Marois2008,Lagrange2010,Macintosh2015,Dupuy2018,Chinchilla2020,Langeveld2024}, 
together with recent advances in opacity calculations \citep{Gharib-Nezhad2021,Chubb2021,Lacy2023}, 
equations of state \citep{Militzer2013, Chabrier2019, Chabrier2021, HG23,TejadaArevalo2024}, 
and interior evolution models involving helium rain and inhomogeneous compositions  \citep{Vazan2016, Muller2020,Knierim2024,TejadaArevalo2025,Sur2025, Knierim2025b,Sur2025c,Sur2025b} requires a new generation of atmosphere models compatible with these 
improved physical inputs.
This motivates us here to update our model grid accordingly. 
Moreover, when used as outer boundary conditions in updated evolutionary calculations, 
this grid will provide a means to tightly couple interior cooling tracks with spectral evolution, 
efficiently leveraging the spectral data already embedded within the atmosphere grids. 

This paper is organized as follows: In \S \ref{sec:model_atmos}, we introduce our numerical methods of computing the static atmospheres, 
highlighting the updated frequency-dependent opacities and the updated equation of state (EOS) tables. In \S \ref{sec:comp_atmosphere}, we compare T-P profiles and spectra generated from our calculation with existing models from ATMOS2020 and Sonora-Bobcat. 
In \S \ref{sec:atmosphere_evolution}, we distill the base entropy information from this set of models into a boundary condition grid with which to evolve interior models. 
In \S \ref{sec:comp_evo}, we compare adiabatic evolution tracks with existing models 
and show next-generation evolutionary tracks with more sophisticated interior physics. 
In \S \ref{sec:interpolator} we present an interpolation toolkit to efficiently post-process evolution tracks into time-resolved spectral evolution. We summarize our work in \S \ref{sec:summary}.

\section{Model Atmospheres: Numerical Setup}
\label{sec:model_atmos}

We generate model atmospheres using the 1D radiative–convective atmosphere code \texttt{CoolTLusty} \citep{Hubeny1995,Sudarsky2003,Sudarsky2005,Burrows2008}. The code computes the T-P structures and emergent spectrum of a plane-parallel atmosphere using an extensive suite of opacity data compiled into thermochemical equilibrium tables \citep{BurrowsSharp1999,SharpBurrows2007}. 
Opacities are treated with line-by-line sampling and have recently been substantially updated by \citet{Lacy2023} (see their Appendix A). 
Their methodology for generating absorption cross sections follows recent advances from the ExoMolOP \citep{Chubb2021} and EXOPLINES \citep{Gharib-Nezhad2021} projects.
To compute absorption cross sections over a grid of temperatures, pressures, and wavenumbers, we use the publicly available Fortran package \texttt{exocross}\footnote{\url{https://exocross.readthedocs.io/en/latest/}}. The code allows user-specified line lists, 
line profiles (typically Voigt, with pressure-dependent broadening if desired), 
line-wing cutoffs, and optional line-strength thresholds. 
For each molecule, 
we adopt the line lists recommended by the ExoMol team, augmented where necessary to extend the high-temperature and broadband coverage.
The full opacity includes line absorption for 15 molecules (H$_2$O, H$_2$, CH$_4$, CO, H$_2$S, NH$_3$, PH$_3$, TiO, VO, SiO, FeH, CaH, CrH, TiH, MgH) and 6 atoms (Li, Na, K, Fe, Rb, Cs), H$^-$ free-free and bound-free absorption, 
and collision-induced absorption (CIA) from both H$_2$-H$_2$ and H$_2$-He. 
Rayleigh scattering is included for H$_2$ and He. The opacity from metals
is assumed to scale linearly with metallicity \citep{SharpBurrows2007}, with relatively ratios determined by solar metallicity $Z_\odot=0.017$. 
For methane, the ExoMol line list of \citet{Yurchenko2014} is supplemented at short wavelengths by cross sections inferred in \citet{Karkoschka1994} from Jupiter’s spectrum and by the associated Rayleigh scattering \citep{Chen2023}.

In our atmospheric structure calculations, we use 100 atmospheric layers and solve radiative transfer at 5000 points evenly spaced in log$_{10}$ wavelength from 0.6 to 300 $\mu$m. 
The primary outputs of our self-consistent calculation are the T-P profile and a 5000-band emergent spectrum, which can subsequently be post-processed into a higher-resolution spectrum with 30000 bands. 
For comparison, 
ATMO2020 computes T-P profiles using 32 correlated-$k$ bands before post-processing to 5000 bands, while Sonora-Bobcat uses 190 correlated-$k$ bands followed by post-processing to 362000 bands. 
For an initial systematic scan over a large parameter space, we neglect disequilibrium chemistry  \citep{HubenyBurrows2007,Lacy2023} or clouds and irradiation \citep{Sudarsky2003,Burrows2006,Burrows2006cloud,Burrows2008,Chen2023}, which will be addressed in future releases. 

For evolutionary calculations of a gas giant in isolation, we emphasize that the atmosphere profile is not only a function of the effective temperature ($T_{\rm eff}$; 
or equivalently the internal flux for isolated planets) and surface gravity $g$, but also the compositional helium and heavy element mass fractions, $Y, Z$.


For helium mixtures ($Y$), we use the H-He EOS tables of \citet{Chabrier2021}, which are equivalent to those of \citet{Chabrier2019} at atmospheric pressures. The H-He EOS of \citet{Chabrier2019} is also applied in the atmosphere calculations of \citet{Marley2021}. We incorporate the ``metal'' EOS of water from \cite{Haldemann2020}, since water is often used to represent heavy elements in interior evolutionary calculations \citep{Vazan2018, TejadaArevalo2025, Sur2025}. 
Past work has incorporated metallicity into atmosphere models \citep[e.g., ATMO2020;][]{Phillips2020}, but the
variations in EOS metallicity were typically accommodated by adopting a H–He EOS with an effective helium fraction $Y' = Y+Z$, where $Z$ is only explicitly considered in the opacity. 
{This approach lacks thermodynamic consistency with the interior evolution and obscures the distinct roles of $Y$ and $Z$} \citep[See][for a more detailed discussion.]{TejadaArevalo2024}. 
{Our method treats metallicity explicitly in the EOS, 
representing a first step toward incorporating heavy elements in a thermodynamically consistent manner. 
While we do not yet model the full set of metals in a way that is fully consistent with the opacity calculations, 
this treatment nevertheless enables a clean separation between the effects of helium fraction and metal abundance.}

\begin{table*}
\centering
\begin{tabular}{lccc}
\hline\hline
 Parameters & Range & Step & Number of Points \\
\hline
$\log_{10} g $& $2.8 \le \log_{10} g \le 4.4$ & 0.2  & 9\\
$T_{\rm eff}$ & $100 \le T_{\rm eff} \le 1400~{\rm K}$  
& 50 ($\le 500$ K), 100 ($>500$ K) & 18\\
$Z$ & $ Z/Z_\odot \in [1, 3.16, 10]$ (vary in both EOS and opacity) & -- & 3\\
$Y$ & $Y \in [0.15, 0.275]$ (vary in EOS) & -- & 2 \\
\hline
\end{tabular}
\caption{Parameter values for our set of atmosphere models.}
\label{tbl:parameter_space}
\end{table*}

To cover the cooling history of 0.3-10 $M_J$ objects from $T_{\rm eff} \sim1500K$ down to $\sim100K$, and their contraction from $\sim 2R_J$ to $\sim R_J$, we choose the range of $T_{\rm eff}$ and $g$ in our atmosphere model grids as indicated in Table \ref{tbl:parameter_space}. 
To self-consistently model atmosphere metallicity growth due to {convective mixing} of elements from the fuzzy core and helium depletion due to helium rain, 
our tables cover three metallicities and two helium fractions, also listed in Table \ref{tbl:parameter_space}. {We do not vary $Y$ in the opacity calculations, 
as its primary effect is to redistribute the relative contributions of Rayleigh scattering and CIA for H$_2$ and He, 
which is subdominant 
compared to the impact of varying metallicity for cold, isolated atmospheres. 
The dependence of CIA opacity on $Y$ may provide constraints 
on the helium content of Jupiter- and Saturn-like irradiated planets with hot tropospheres \citep{Gautier1981,Achterberg2020}, and will be accounted for 
in future studies that incorporate irradiation. }
The total number of atmospheric models is $9\times 18\times 3\times 2 = 972$. 
{Note that our model grid covers a fiducial helium fraction $Y = 0.275$, 
which is slightly higher than the value used in SB21 ($Y = 0.2735$) and consistent with ATMO2020. We also adopt a solar metallicity normalization of $Z_\odot = 0.017$, 
compared to $Z_\odot = 0.0153$ in SB21 and $Z_\odot = 0.0169$ in ATMO2020. 
These small differences should not significantly affect the comparisons presented below.}

\section{Comparison of Atmospheric Models}
\label{sec:comp_atmosphere}
\subsection{Thermal Profiles}

\begin{figure*}[htbp!]
\centering
\includegraphics[width=0.45\textwidth,clip=true]{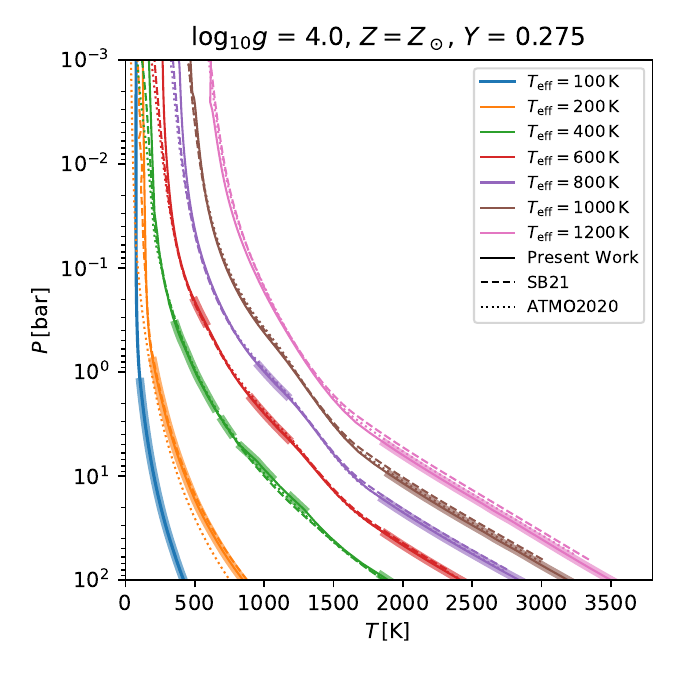}
\includegraphics[width=0.45\textwidth,clip=true]{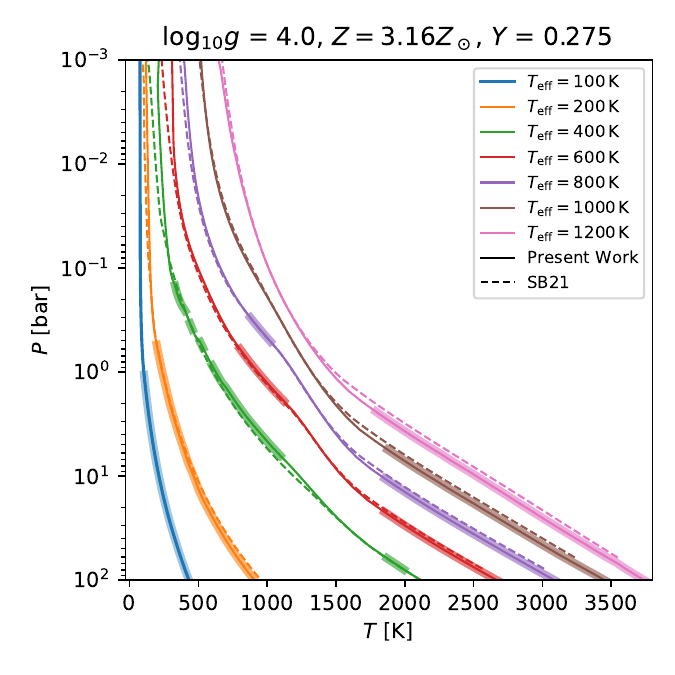}
\caption{Radiative–convective equilibrium temperature–pressure (TP) profiles from our atmosphere models compared with those at a representative surface gravity $\log_{10} g = 4.0$ and $Y = 0.275$. 
The left panel shows solar-metallicity models compared with ATMO2020 (dotted) and Sonora-Bobcat (SB21; dashed) \citep{Phillips2020, Marley2021}, and the right panel shows enhanced-metallicity models compared with SB21. 
Our T-P profiles (solid lines) generally agree with those of ATMO2020 and SB21 in the deeper atmosphere {($P \gtrsim 1$ bar where the base entropy is determined)}, with the largest differences occurring at 200 K. 
The thick line segments identify the convective regions in our T-P profiles.}
 \label{fig:thermal}
\end{figure*}

Figure \ref{fig:thermal} presents radiative–convective equilibrium T-P profiles for a representative subset of our models, 
shown alongside profiles from existing atmospheric grids. In each panel, this work (solid lines) is compared with the Sonora–Bobcat \citep[SB21][]{Marley2021} models (dashed lines).
The left panel displays T-P profiles at solar metallicity, with ATMO2020 \citep{Phillips2020} profiles overlaid (dotted lines).
The right panel shows T-P profiles at {3.16 solar metallicity, motivated by measured and inferred water abundances of Jupiter \citep{LiCheng2020,Hyder2025}}.
In both panels, the thick line segments show convective regions in our T-P profiles.
The profiles from different papers generally agree, 
including in the existence of detached convective zones within the range of effective temperatures 600-800K \citep{Burrows1997,Marley2021}. 
{The temperature differences in the deeper atmosphere ($P \gtrsim 1$ bar, where the profiles approach the convective region and set the interior entropy) are typically within $\sim$5\% relative to ATMO2020 and $\sim$2\% relative to SB21. 
Larger deviations can occur near the upper boundary, but they do not affect the cooling history.}


\subsection{Model Spectra}

\begin{figure*}[htbp!]
\centering
\includegraphics[width=0.49\textwidth,clip=true]{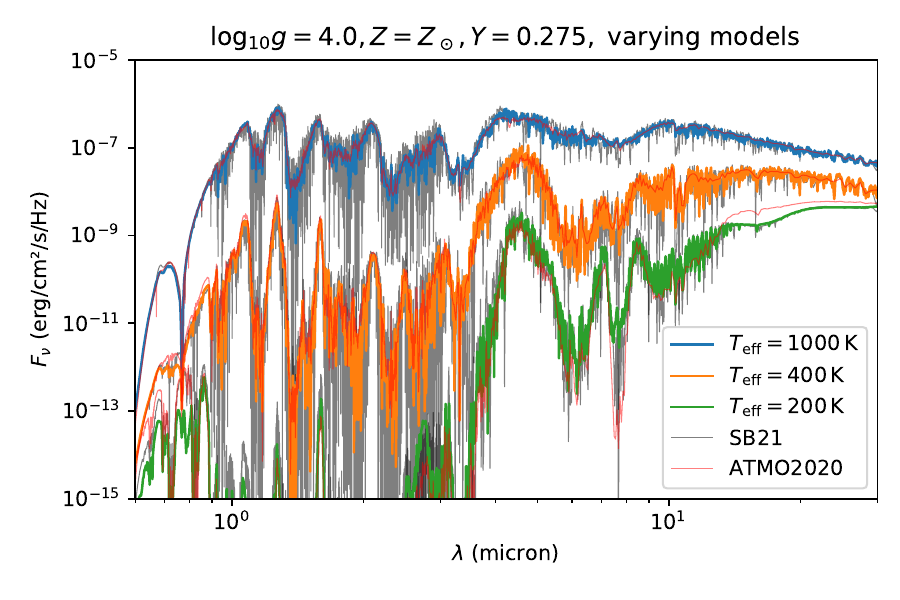}
\includegraphics[width=0.49\textwidth,clip=true]{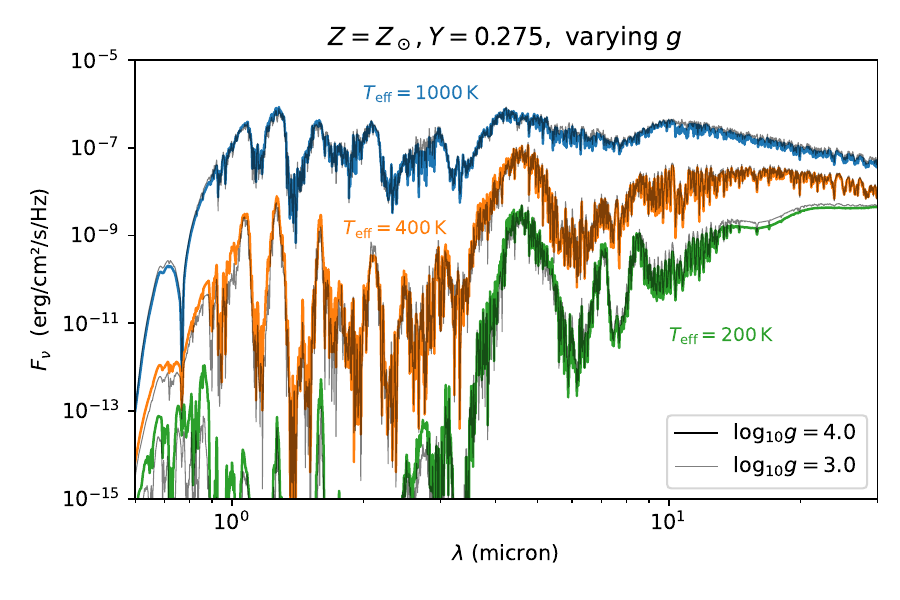}
\includegraphics[width=0.49\textwidth,clip=true]{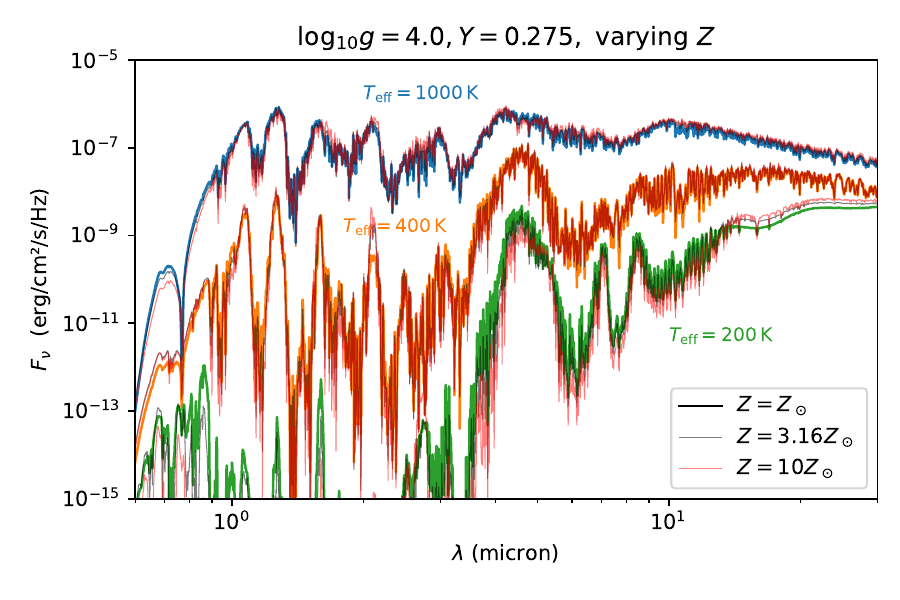}
\includegraphics[width=0.49\textwidth,clip=true]{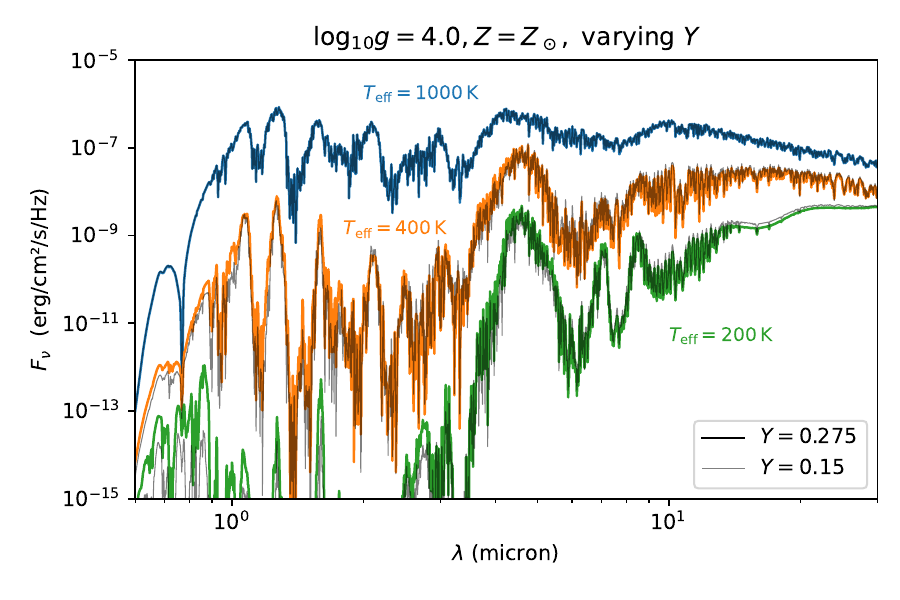}
\caption{Comparison of spectra from different atmospheric models. 
Upper left: Spectra from our models with log $g = 4.0$, $Z = Z_\odot, Y = 0.275$, 
shown alongside spectra from the Sonora-Bobcat \citep[SB21][]{Marley2021} and ATMO2020 \citep{Phillips2020} grids at the same gravity and metallicity. 
Upper right: Our models' dependence on varying $\log_{10} g = 3.0, 4.0$. 
{Lower left: Our models' dependence on varying metallicity $Z = Z_\odot, 3.16Z_\odot, 10Z_\odot$. 
Lower right: Our models' dependence on varying helium abundance $Y=0.15, 0.275$.} Note that all the blue, orange, and green solid lines across the four panels are identical for comparison. Our spectra generally agree with SB21 and are most sensitive to changes in $Z $. }
 \label{fig:spectra}
\end{figure*}

\begin{figure*}[htbp!]
\centering
\includegraphics[width=0.98\textwidth,clip=true]{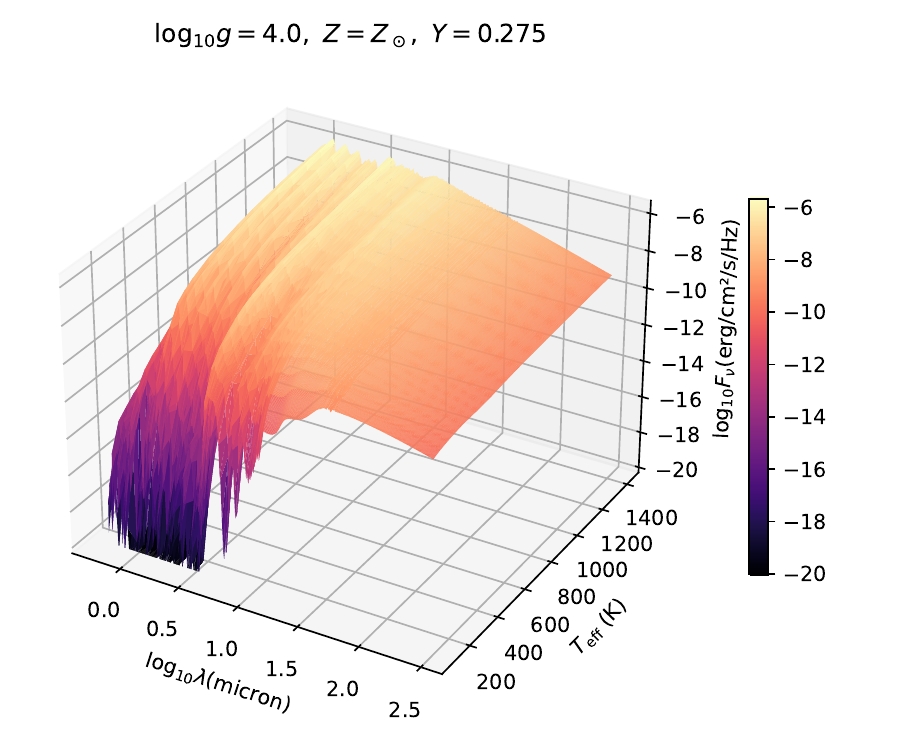}
\caption{Spectra over a range of effective temperatures presented as 3D surfaces for a typical $(\log_{10} g, Z, Y)$-parameter combination. 
The rapid suppression of sub-micron flux at low effective temperatures is particularly pronounced 
and will be manifest in the time evolution of the spectrum (see \S \ref{sec:interpolator}). }
 \label{fig:spectrum_3D}
\end{figure*}

We provide in Figure \ref{fig:spectra} a comparison of spectra for different atmosphere models. 
The upper left panel depicts spectra for $T_{\rm eff} = 200, 400, {\ \rm and\ }1000$~K from our log$_{10}$ $g = 4.0$, $Z = Z_\odot, Y = 0.275$ models, 
as well as a comparison with log$_{10}$ $g = 4.0$, $Z = Z_\odot$ models in the Sonora-Bobcat and ATMO2020 catalogs (shown in grey and red respectively). 
The upper right panels depict spectra for our models at different gravities (log$_{10}$ $g = 3.0,4.0$), the lower left panel compares different metallicities ($Z = 1, 3.16, 10 Z_\odot$), and the lower right panel compares two values of $Y$ (0.15, 0.275). All panels compare the same fiducial effective temperatures. 
The spectral shapes are generally insensitive to these parameters, 
although decreasing $g$ and increasing $Z$ suppress flux at short wavelengths ($\lesssim 3$ micron) and enhances emission at longer wavelengths by $\sim$0.1 dex. 
This behavior is most pronounced at lower effective temperatures, 
where molecular opacity dominates absorption.
An increase in $Z$ directly enhances molecular opacities by raising the abundance of molecular absorbers. 
In contrast, a decrease in surface gravity primarily acts by shifting the atmosphere to lower pressures at fixed optical depth, 
which favors molecule formation. 
Generally, enhanced molecular opacity preferentially attenuates the emergent flux $\lesssim 5\ \mu$m, 
with the absorbed energy redistributed and re-emitted at $\gtrsim 5 \ \mu$m, 
although there is the significant outlier of $K$-band water absorption feature that increases in flux at higher $Z$, lower $Y$ and lower $g$, 
as also highlighted in \citet{Marley2021}. 
For completeness, we also show the 3-D presentations of our spectral models over the entire effective temperature range for the fiducial parameter combinations in Figure \ref{fig:spectrum_3D}. The full dataset is used to generate spectral evolution histories in \S \ref{sec:interpolator}.

\section{Atmosphere Models as Boundary Conditions For Evolution}
\label{sec:atmosphere_evolution}
\begin{figure*}[htbp!]
\centering
\includegraphics[width=0.46\textwidth,clip=true]{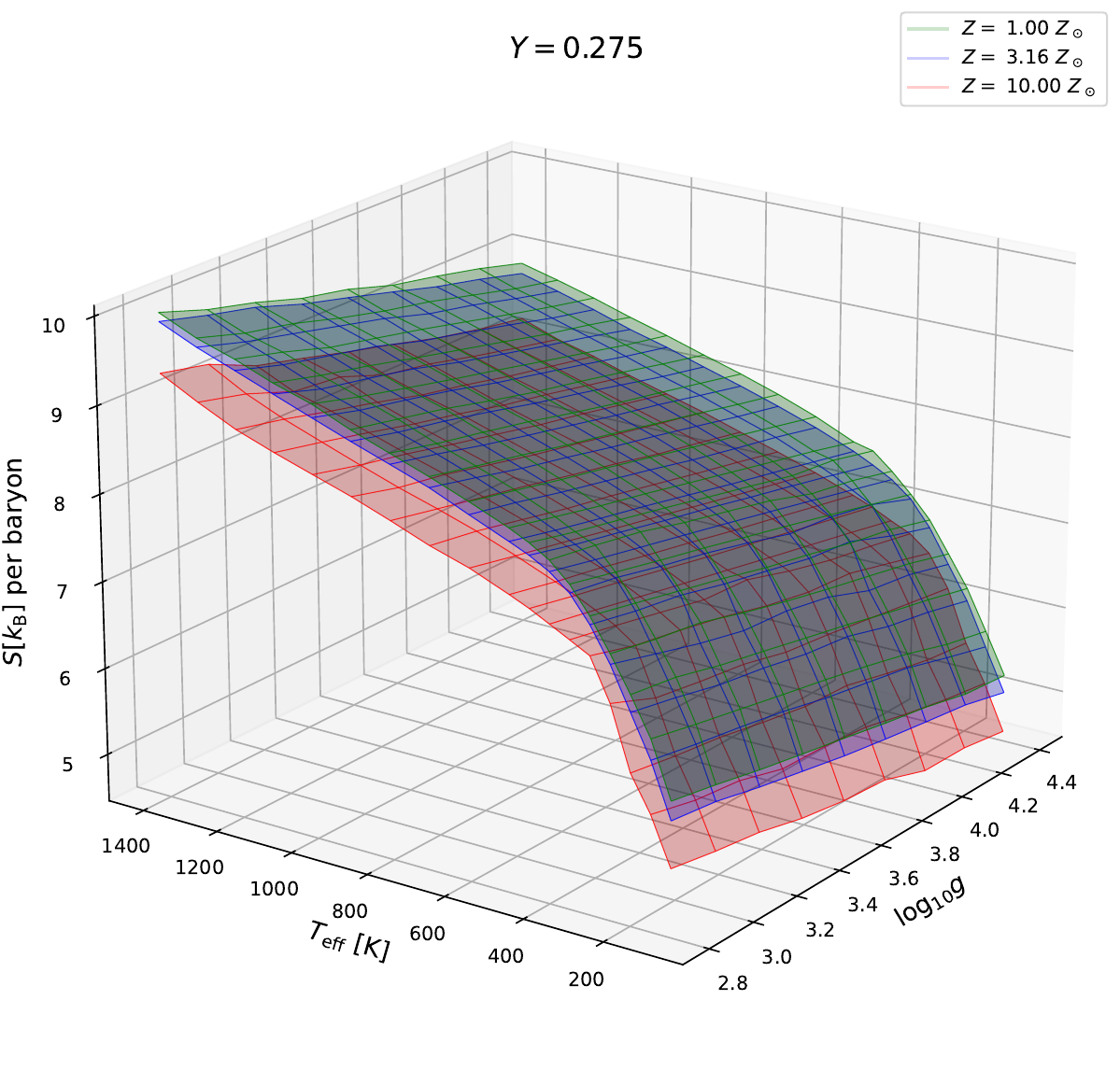}
\includegraphics[width=0.46\textwidth,clip=true]{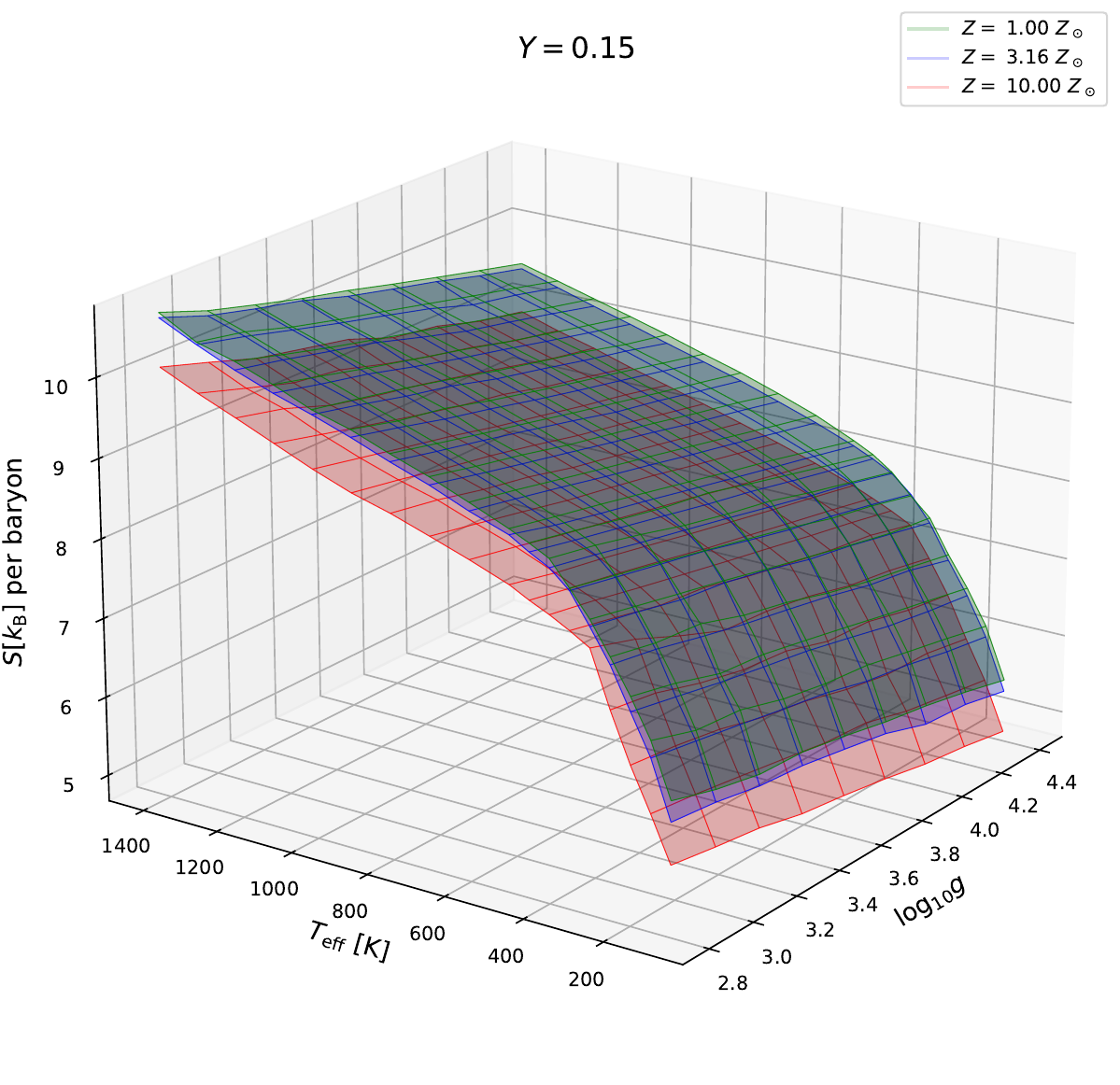}
\caption{$S(T_{\rm eff}, \log_{10} g)$ surfaces at 3 different metallicities and for $Y = 0.275$ (left) and $Y = 0.15$ (right). The entropy change between $Z =1, 3.16 Z_\odot$ (green and blue sheets, nearly overlapping) is relatively modest, 
while the $10 Z_\odot$ table (red) has significantly lower relative entropy by about 0.7 k$_B$ baryon$^{-1}$. This implies that higher metallicity atmospheres accelerate cooling at early ages. {A decrease in $Y$ similarly shifts the entropy surface to lower values, 
altering the mapping between interior entropy and $T{\rm eff}$ and thereby providing feedback on the potential heating effect of helium rain (see \S \ref{sec:comp_evo}). }}
 \label{fig:entropy_surface}
\end{figure*}

Each radiative–convective equilibrium profile maps to an entropy value $S$ below the deep radiative–convective boundary in the planet's outer convective zone. 
The relationship between entropy, 
effective temperature and surface gravity (at given values of $Y$ and $Z$) ultimately govern the planet's thermal evolution. 
This information can be encoded with an entropy table in a $S(T_{\rm eff}, \log_{10} g, Y, Z)$ four-dimensional parameter space.  
In the form of overlaid three-dimensional surfaces, 
Figure \ref{fig:entropy_surface} provides a dimensionally-reduced illustration of a table for $S(T_{\rm eff}, \log_{10} g)$ at $Y = 0.275$ (left panel) and $Y = 0.125$ (right panel), 
where each sheet represents three different values of $Z$. These sheets show that, 
at the same effective temperature, higher metallicity yields lower entropy at the base of the atmoshere. 
{This is consistent with the general thermodynamic trend that
heavier elements have lower entropy than lighter elements for the same temperature. 
The largest entropy differences at the same effective temperature are seen between 3.16 and 10 times solar on both panels, 
constituting an entropy difference of $\sim$0.7 k$_B$ baryon$^{-1}$. }

By inverting this entropy table, one can express 
$T_{\rm eff}$ as a function of surface gravity and the deep interior entropy $S$. 
The latter is sometimes parameterized by $T_{10}$, the temperature at 10 bars, 
assuming that this level lies within the deep convective zone \citep{Burrows1997,Hubbard1999,Fortney2011}. 
This relation enables the construction of evolutionary tracks \citep[e.g.,][]{Burrows1997, Fortney2004, Fortney2007, Marley2021, Sur2025b} by self-consistently matching an interior model to its emergent flux. 
In the evolutionary calculations presented in \S \ref{sec:comp_evo}, we adopt a high initial entropy (``hot-start”) configuration 
and compute cooling tracks using both 
(i) a standard adiabatic interior model \citep{Fortney2007,Saumon2008,Marley2021} (for comparison with existing literature) and 
(ii) a composition-dependent model  \citep[e.g.][]{TejadaArevalo2025, Sur2025b, Sur2025, Sur2025c} with inhomogeneous, non-adiabatic evolution incorporating treatments of fuzzy core mixing and/or helium rain.


\section{Comparison of Evolutionary Tracks}
\label{sec:comp_evo}
\begin{figure*}[htbp!]
\centering
\includegraphics[width=0.9\textwidth,clip=true]{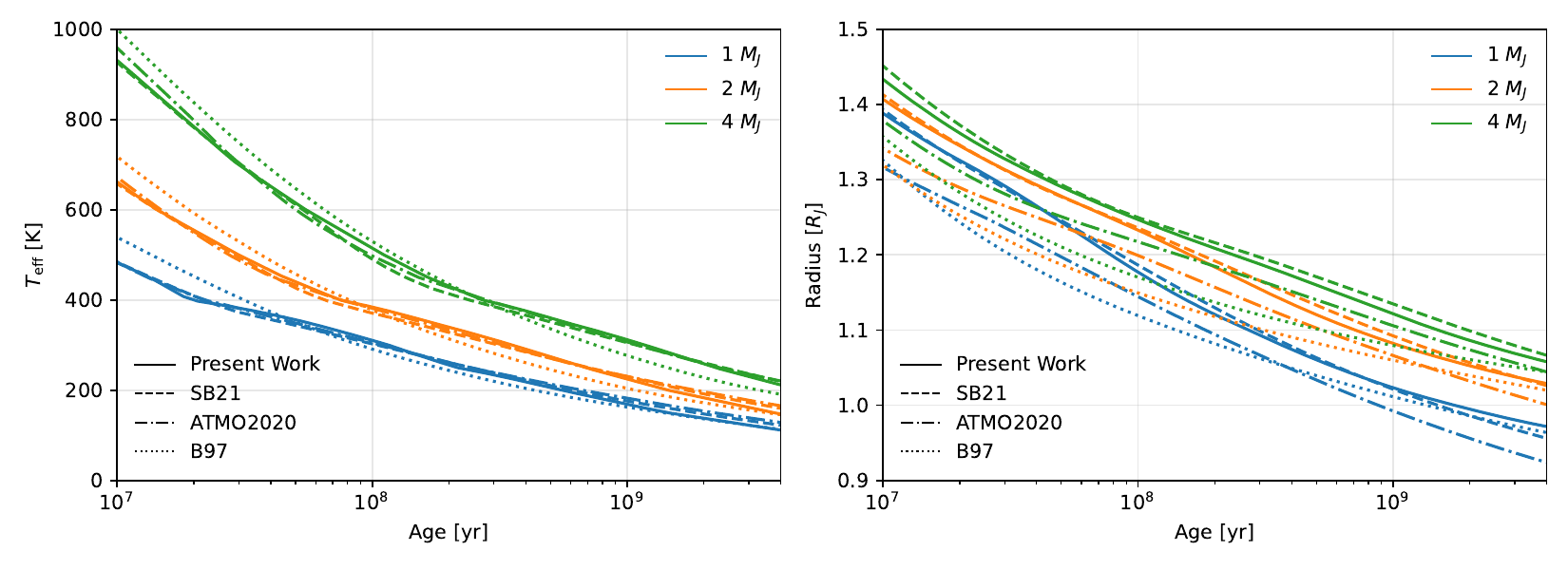}
\caption{Adiabatic evolution of giant planets with different masses (different colors) with fixed and homogeneous compositions $Z = Z_\odot, Y = 0.275$. 
{Different atmosphere models are shown in different line styles. 
In terms of $T_{\rm eff}$, our evolution tracks are similar to those of SB21 and ATMO2020, 
while B97 is hotter at early times but cools more rapidly and becomes colder at late times ($\gtrsim$ a few $\times 10^8$ yr). 
In terms of radius, 
our models closely follow SB21 and are generally larger than ATMO2020 and B97 by $\sim$10\%. For the $1,M_J$ case, the B97 radius converges to our model at late times ($\sim 0.96,R_J$), whereas ATMO2020 remains more compact ($\sim 0.92,R_J$).}}
 \label{fig:adiabatic_evolution}
\end{figure*}

We use our gas giant evolution code, \texttt{APPLE} \citep{Sur2024}, with its flexibility to incorporate any general boundary condition, to generate adiabatic evolutionary tracks of 1, 2, and 4 M$_J$ gas-giant planets using the SB21, 
ATMO2020 boundary as well as \citet{Burrows1997} boundary conditions. 
{Although these works provide their own tabulated evolutionary tracks, 
our results are recomputed with our own interior solver to enable a controlled comparison across boundary conditions. 
We also compute our fiducial tracks using our own boundary condition for fixed composition
$Z = Z\odot, Y = 0.275$, and plot the resulting cooling histories (effective temperature and radius as functions of time) in Figure~\ref{fig:adiabatic_evolution}.}
Colors show different masses, 
and the line styles indicate boundary conditions from different works. 
{In terms of $T_{\rm eff}$, 
our models closely track SB21 and ATMO2020, 
whereas B97 remains hotter at early times, but cools more rapidly and becomes colder at later stages ($\gtrsim$ a few $\times 10^8$ yr). 
For the radius evolution, 
our results are similar to SB21 and generally produce larger radii than B97 and ATMO2020 by $\sim$10\%, particularly for higher-mass planets, although the differences are smaller for the $1,M_J$ case at late times.}

As we emphasized, the existing conventions for homogeneous and adiabatic calculations are gradually being updated by models that incorporate more realistic, sophisticated interior physics 
\citep{Vazan2018, TejadaArevalo2025, Knierim2025b, Sur2025b}. 
In particular, a non-trivial initial fuzzy/dilute core metallicity profile, 
{convective mixing of elements}, and helium rain, 
once included, 
can significantly affect the atmospheric $Y$ and $Z$ values. 
Such forward modeling motivates us to prepare boundary-condition tables for $S$ over the 4-D parameter space $(T_{\rm eff},\log_{10} g, Y, Z)$. {While calculating the 1-D evolution models, we obtain accurate atmosphere radiative fluxes using not only $S,\log_{10} g$, but also the $Y, Z$ values of the outermost mass shell.
Henceforth, we denote these surface values as $Y_{\rm atm}$ and $Z_{\rm atm}$ to distinguish them from the internal helium and metallicity profiles.}

\begin{figure*}[htbp!]
\centering
\includegraphics[width=0.9\textwidth,clip=true]{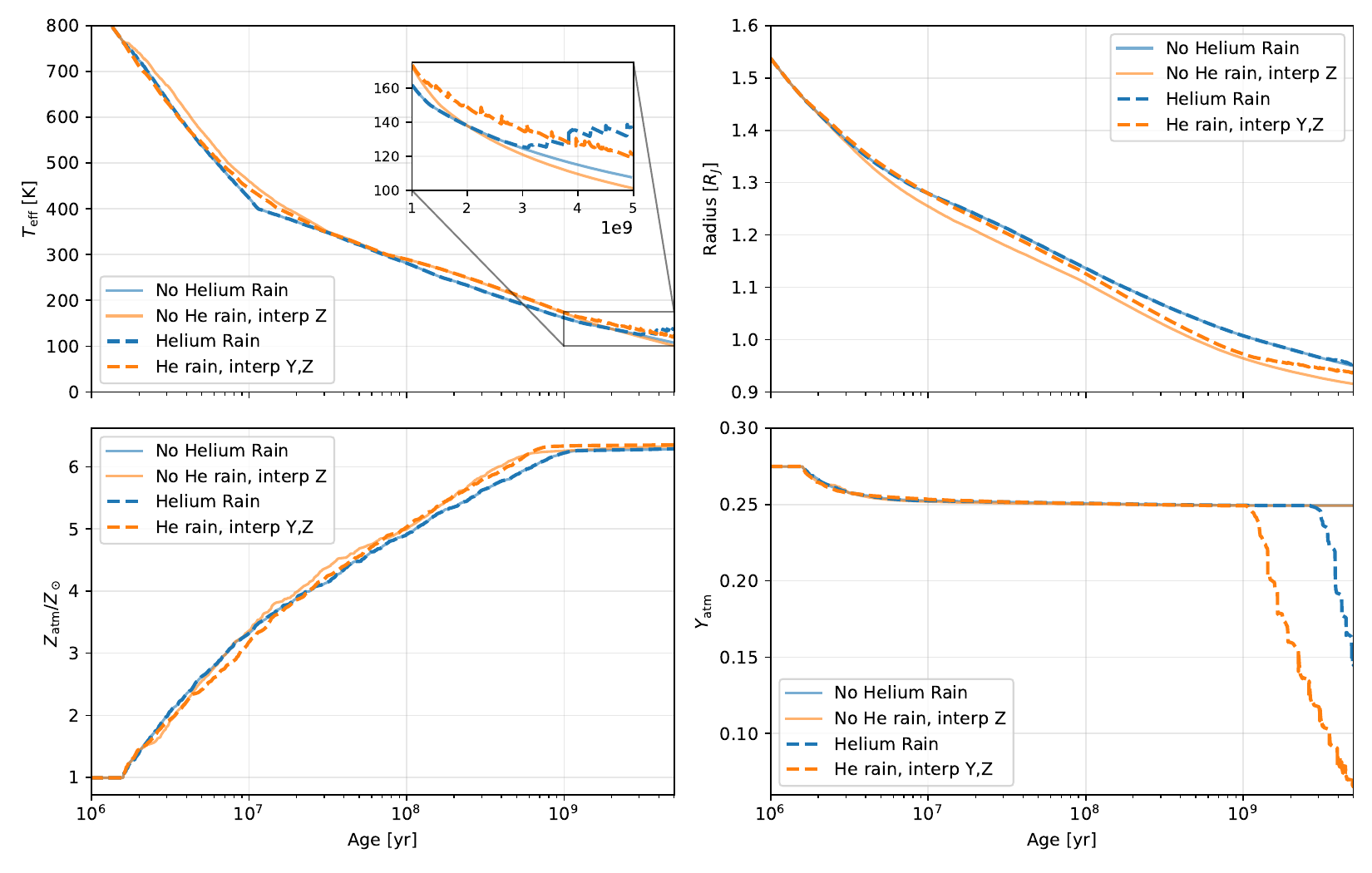}
\caption{Non-adiabatic and compositionally inhomogeneous evolution of 1 $M_J$ models initialized with fuzzy, stably stratified cores.
Top left: effective temperature $T_{\rm eff}$ (with linear-time insets highlighting small-scale features).
Top right: radius evolution.
Bottom left: atmospheric metallicity $Z_{\rm atm}/Z_\odot$.
Bottom right: atmospheric helium abundance $Y_{\rm atm}$.
Dashed curves include helium rain when interior temperatures fall below the helium immiscibility threshold 
\citep{Lorenzen2009, Lorenzen2011}, while semi-transparent solid curves neglect helium rain. 
{Dashed blue lines use atmosphere boundary conditions that neglect the evolution of atmospheric composition, 
holding $Y_{\rm atm}$ and $Z_{\rm atm}$ fixed at their initial values, whereas the dashed orange lines use model atmospheres whose opacities and structure self-consistently respond to the evolving $Z_{\rm atm}$ or both $Y_{\rm atm}$ and $Z_{\rm atm}$ (interpolating across surfaces in Figure \ref{fig:entropy_surface}).
All models begin with $Z_{\rm atm}=Z_\odot$ and $Y_{\rm atm}=0.275$. 
Allowing the atmospheric opacity and structure to respond to evolving $Y_{\rm atm}$ and $Z_{\rm atm}$ cools the planet more rapidly, causing the onset of helium rain to occur sooner by $\sim$2 Gyr compared to models without interpolation.} As a result, the dashed orange model 
($Y_{\rm atm}, Z_{\rm atm}$ interpolation) experiences more severe helium depletion compared to the blue model 
(no $Y_{\rm atm}, Z_{\rm atm}$ interpolation). 
The inclusion of self-consistent metallicity boundary conditions in the interior can therefore affect current evolutionary models of Jupiter and Saturn, as well as exoplanets in general.}
 \label{fig:non-adiabatic_evolution}
\end{figure*}

To illustrate these effects, Figure~\ref{fig:non-adiabatic_evolution} shows the evolutionary tracks of $1\ M_J$ models initialized with fuzzy, stably stratified cores with $Y$ and $Z$ gradients. 
Two models (dashed lines) include helium rain once the interior temperature drops below the helium immiscibility threshold from \citet{Lorenzen2009,Lorenzen2011} \footnote{{We note that the location of this threshold is uncertain. Homogeneous interior models generally favor a threshold shifted to lower temperatures \citep{Nettelmann2015,Mankovich2016,Howard2024}, 
whereas models that include inhomogeneous evolution with helium rain \citep{Sur2025,TejadaArevalo2025}, as well as recent experimental constraints \citep{Brygoo2021}, 
suggest that the boundary may lie at higher temperatures. Here we adopt the unshifted values.}}. 
All models begin with an atmospheric composition, $Z_{\rm atm}=Z_\odot$ and $Y_{\rm atm}=0.275$. 
{Because the initial metallicity profile is prescribed, 
the calculation is not intended as a fit to Jupiter (which requires more adjustments of initial conditions and inclusion of irradiation), 
but rather as a controlled comparison of different atmosphere boundary treatments  relevant for gas giants in general. 
For example, the strong helium depletion in some models exceeds Jupiter’s observed value, but is more comparable to that of Saturn.}
We compare those against the standard treatment of atmospheric boundary conditions, 
in which interpolation is not performed and $Y_{\rm atm}$ and $Z_{\rm atm}$ are held fixed at their initial values\footnote{Effectively, 
this corresponds to interpolating only along one $Z = Z_\odot$ surface shown in Figure~\ref{fig:non-adiabatic_evolution}, which is common practice in literature without a boundary condition that covers the composition parameter space.} (blue curves). 
We contrast this with models that interpolate in metallicity, 
or in both helium fraction and metallicity (orange curves), thereby accounting for the physical effects of an evolving atmospheric composition.

To isolate the effect of metallicity-dependence alone, first we compare the semi-transparent solid blue and orange curves, 
which differ only in whether the model atmospheres account for variations in $Z_{\rm atm}$ (at fixed $Y_{\rm atm}$). 
Since the initial entropy profiles and radii are identical, the initial divergence in $T_{\rm eff}$ arises purely from how the atmospheric boundary condition is interpolated. 
At $Z_{\rm atm}>Z_\odot$, 
the entropy at fixed $g$ and $T_{\rm eff}$ is smaller than at $Z_{\odot}$ (see Figure \ref{fig:non-adiabatic_evolution}). 
Consequently, 
for a given interior entropy and gravity, a more metal-rich atmosphere corresponds to a higher inferred $T_{\rm eff}$. 
This explains the initially higher $T_{\rm eff}$ in the models that allow the atmospheric structure to respond to increasing $Z_{\rm atm}$ when their temperature evolutions first diverge at $\sim 10^6$ yr (see top-left panel of Figure~\ref{fig:non-adiabatic_evolution}).
The change in metallicity arises from convective mixing, 
which gradually erodes any primordial compositional gradient over evolutionary timescales. 
This redistribution 
gradually enriches the atmosphere, increasing $Z_{\rm atm}$ to $\sim 5$–$7 Z_\odot$ (see lower left panel of Figure~\ref{fig:non-adiabatic_evolution}) and smoothing $Y_{\rm atm}$ towards a value of $\sim$ 0.25 even without the onset of helium rain. {For the faint blue line, 
the sharp local curvature of the $S(T_{\rm eff}, Z_{\rm atm}=Z_\odot)$ relation at $T_{\rm eff}\sim 400$K 
(see Figure \ref{fig:entropy_surface}) produces a small kink in the cooling track. Using finer atmosphere grids would dampen this discontinuity. Interestingly, in models where the atmospheric structure responds to evolving $Z_{\rm atm}$  (faint orange line), 
the effect is already largely smoothed out because the evolution samples a range of neighboring $S(T_{\rm eff})$ relations.}

Nevertheless, these differences is subsequently regulated by feedback. 
A larger $T_{\rm eff}$ enhances radiative cooling, 
accelerating entropy loss 
and gradually diminishing the difference between the cooling tracks. 
At late stages, the effective temperatures tend to converge because combinations of higher metallicity and lower entropy map to similar locations in $(T_{\rm eff}, \log_{10} g)$ (or equivalently $(T_{\rm eff}, R)$) space. 
As a result, 
direct observables such as $T_{\rm eff}$ and radius become hard to distinguish even 
though the more rapid cooling leads to a $\lesssim$ 5\% more compact planet radius in the interpolated model. 
Apart from these, the $Y_{\rm atm}$ and $Z_{\rm atm}$ changes through {convective mixing} are nearly identical.

However, we find that differences in both $T_{\rm eff}$ and $Y_{\rm atm}$ evolution
become significant once additional physics, 
particularly helium rain, 
is taken into account.
In Figure~\ref{fig:non-adiabatic_evolution}, we also plot the helium rain model without interpolation (dashed blue lines), 
whose evolution is identical to the semi-transparent solid blue lines 
until helium rain is initiated at $\sim 3$ Gyr, and $Y_{\rm atm}$ sharply decreases. 
Note that since we adopt the classical Schwarzschild criterion for convection here, 
helium rain heats the outer layers \citep{Fortney2003}, causing an increase in effective temperature. Using $Y_{\rm atm},Z_{\rm atm}$ leads to slower cooling or even an increase in $T_{\rm eff}$ seen in the right inset of the top-left panel of Figure~\ref{fig:non-adiabatic_evolution}, 
where the semi-transparent and dashed curves diverge significantly once helium rain sets in. 
Under the Ledoux criterion, helium rain instead cools the outer layers \citep{Mankovich2016, Sur2025c}. 
Which prescription is more appropriate remains an open question and is not the focus of this paper. 

Finally, the model that includes helium rain and consider the variation of both $Y_{\rm atm}$ and $Z_{\rm atm}$ is shown by the dashed orange curves. 
As in the semi-transparent orange case, 
variation in $Z_{\rm atm}$ initially produces a higher $T_{\rm eff}$ than the blue curves prior to significant changes in $Y_{\rm atm}$, 
leading to a faster decrease in entropy and faster cooling of the interior temperature profiles. 
Consequently, helium rain is triggered sooner at $\sim 1$ Gyr, accelerating the depletion of atmospheric helium. 

{It is worth noting that once helium rain begins, allowing the atmospheric boundary conditions to respond to the evolving $Y_{\rm atm}$ leads to a significantly milder suppression of cooling compared to the more pronounced $T_{\rm eff}$ inversion seen in the helium rain model with fixed atmospheric composition (blue dashed line). 
In the latter case, helium rain heats the envelope while the atmospheric composition remains fixed, 
so the mapping from increasing entropy to $(T_{\rm eff}, g)$ along a single boundary surface produces a rapid increase in $T_{\rm eff}$ (see Figure~\ref{fig:entropy_surface}).
By contrast, when the atmosphere composition self-consistently reflects the depletion of helium, 
the boundary condition shifts such that, 
at a given entropy and similar surface gravity, lower $Y_{\rm atm}$ corresponds to a lower $T_{\rm eff}$. 
As a result, the heating effect of helium rain is partially offset, yielding a more self-limiting and internally consistent evolution.
By $\sim$5 Gyr, 
these differences lead to substantial deviations in observable properties. The atmospheric helium fraction is reduced by $\sim$0.1 relative to the helium rain model with fixed composition, and by $\sim$0.2 compared to models without helium rain. Meanwhile, 
the effective temperature and radius lie between the two limiting behaviors: 1) the cooler, more compact evolution without helium rain, 
and 2) the hotter, more inflated evolution with helium rain, but fixed atmospheric composition.
This highlights the importance of treating atmospheric composition and helium rain in a unified, self-consistent framework. 
In future work, we will adopt this approach to explore a broader parameter space of planetary masses, initial entropies, and compositional profiles, while systematically varying key processes such as convective mixing and helium rain.}

\section{Interpolator for Spectral Evolution}
\label{sec:interpolator}
To better connect our evolutionary models with observations, 
it is necessary to generate spectra at arbitrary points along continuous evolution tracks. 
Directly imaged cold Jupiters and brown dwarfs certainly 
do not coincide with discrete grid points in $(T_{\rm eff}, \log_{10} g, Y, Z)$ space, 
and atmospheric retrieval frameworks similarly require spectra evaluated at continuously varying parameters. 
We therefore construct 
an efficient interpolator that enables spectral predictions along evolutionary trajectories and at arbitrary locations in our 4D parameter space.

Given the piecewise-uniform structure of our tabulated grid, 
the first interpolation step, 
namely, identifying the bracketing grid points in each dimension, 
can be implemented efficiently using simple integer arithmetic rather than generic search routines. 
In particular, 
the grid indices corresponding to a target parameter set can be obtained via direct arithmetic indexing within each grid segment, 
eliminating the need for binary searches. This reduces the computational complexity of index lookup from the usual $\mathcal{O}(\log N)$ complexity to $\mathcal{O}(1)$ where $N$ is the characteristic length of each dimension. 
Once the neighboring grid points are identified, the spectrum (flux) at the desired parameter location is obtained by multilinear interpolation, 
combining the spectra at the adjacent grid points with weights determined by the relative offsets from neighboring points.  
A script implementing this interpolation scheme is publicly available, allowing users to generate spectra at specified locations or along trajectories in the $(T_{\rm eff}, \log_{10} g, Y, Z)$ parameter space. 
In addition, the associated spectra will enable computation of the photometric evolution in arbitrary bands or user-specified filters. 

\begin{figure*}[htbp!]
\centering
\includegraphics[width=1\textwidth,clip=true]{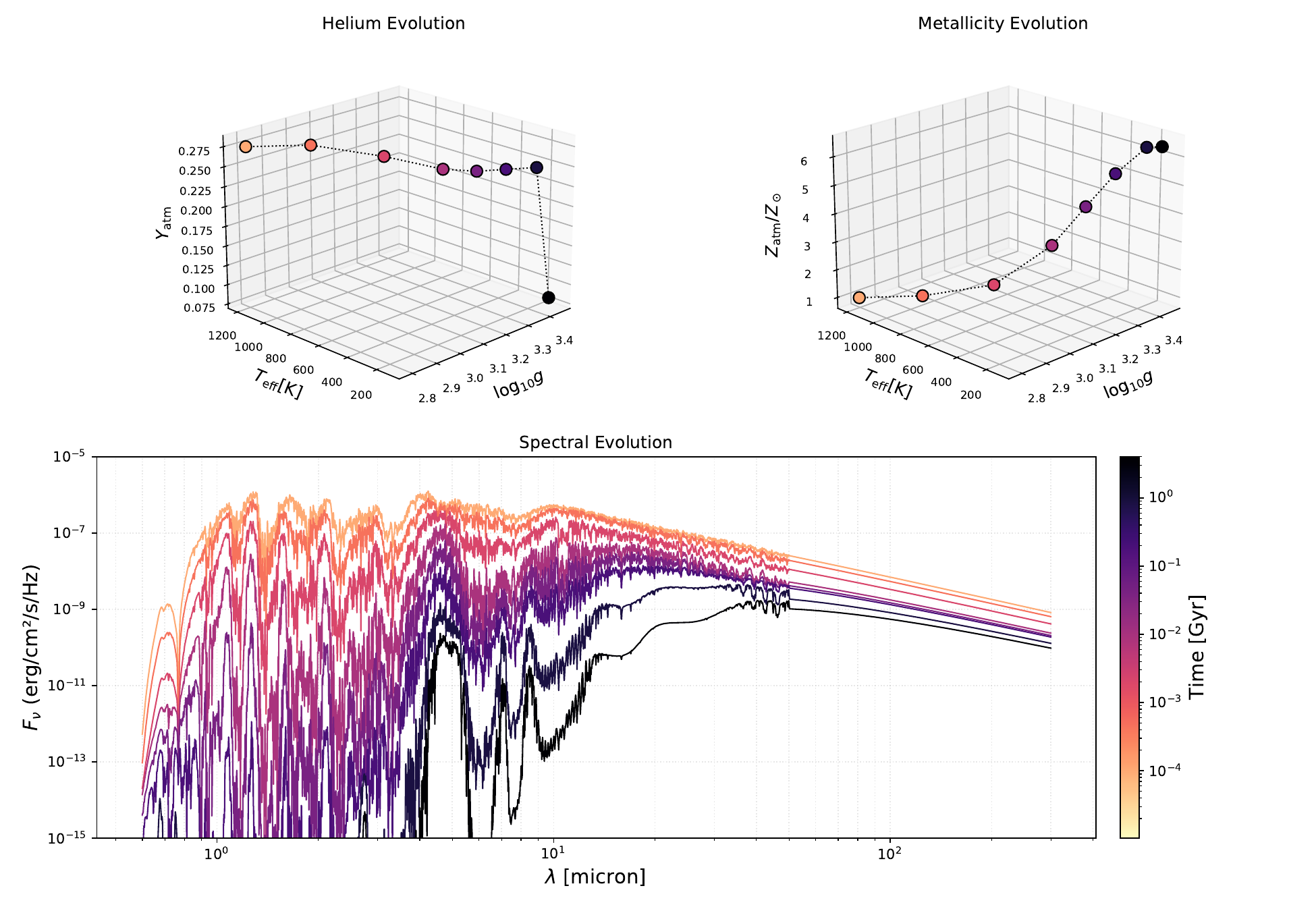}
\caption{Top panels: the evolutionary trajectory of the atmosphere parameters $Y$ and $Z$
in the model with convective mixing of a fuzzy core and helium rain in the $T_{\rm eff}, g$ parameter space. Colored points mark discrete snapshots sampled along the evolution, with color indicating time (logarithmically scaled; see colorbar). 
Bottom panel: the corresponding interpolated spectral evolution obtained by post-processing the trajectory using the atmospheric grid. 
For each time step, the emergent spectrum is computed via multilinear interpolation $T_{\rm eff}, \log_{10} g$ as well as  composition parameters. This illustrates how the interpolator maps a thermodynamic and compositional evolution track into a continuous sequence of spectra.}
 \label{fig:interpolator}
\end{figure*}

As an illustration, 
Figure \ref{fig:interpolator} shows in its lower panel the 
spectral evolution along the evolutionary track 
of the ``complete" model with helium rain as well as the associated interpolation in atmospheric $Y, Z$ (dashed orange lines in Figure \ref{fig:non-adiabatic_evolution}) 
at different times. 
In the top panels, 
points along the trajectory 
are plotted as colored markers on the corresponding $(T_{\rm eff} \log_{10} g, Y)$ or $(T_{\rm eff}, \log_{10} g, Z)$ parameter space, where the beads represent different snapshots corresponding to each spectrum.





\section{Summary} \label{sec:summary}

In this study, we have developed an extensive suite of atmospheric models using the 1D atmosphere and spectral code \texttt{CoolTLusty}. 
Our goal was to produce state-of-the-art boundary condition tables suitable for modeling the evolution of gas giant planets with masses $0.3-10M_J$. 
Our models incorporate substantially updated atmospheric opacities \citep{Lacy2023} and an improved equation of state that accounts for independent variations in helium fraction and metallicity \citep{Chabrier2021, Haldemann2020, TejadaArevalo2024}. 
The resulting grid spans a four-dimensional parameter space, 
and we benchmark our temperature–pressure profiles and spectra against those from previous model sets.

These tables enable the construction of evolutionary models for the cooling and contraction of gas giants 
that allow for both metallicity and helium fraction changes due to compositional mixing and helium rain. 
To facilitate comparison with existing work, we computed 
evolutionary tracks assuming adiabatic interiors using the interior evolution code \texttt{APPLE}, 
demonstrating consistency with results obtained from traditional boundary-condition tables. 
Additionally, we presented sample tracks derived from this more sophisticated non-adiabatic interior-evolution framework, which includes convective mixing and helium rain. 
Finally, 
we developed an efficient interpolator that generates time-resolved spectral evolution along arbitrary trajectories in $(T_{\rm eff}, g, Y, Z)$ space, 
allowing direct prediction of observables such as luminosity, 
colors, 
and band-limited light curves throughout a planet’s time evolution continuously across the atmosphere parameter space. 

This new boundary-condition dataset, 
as well as the spectrum interpolator tool,
is available to the scientific community for future research into the evolution and spectral appearance of giant exoplanets at \href{https://zenodo.org/records/18832237?preview=1&token=eyJhbGciOiJIUzUxMiJ9.eyJpZCI6IjUzZDUwMDg5LWEzMmUtNDA1NC1hY2EzLTY0ODM2N2Y2ZjQ2YiIsImRhdGEiOnt9LCJyYW5kb20iOiIwZWM5MmFmMDgxYjg0NTZkYWMyOTJlNTVhNmJhMGVhNSJ9.SJzNTSibv_p7mU40zK5PYApwaUBJLhMPHJYeP0AC0mhl5_3VaGwroZRwn0HwkiMt5JbUtoKmWwrAKEqQpVIVBA}{this link}. 
In this release, 
we consider cloudless, isolated giant planets in chemical equilibrium. 
Tables incorporating clouds, 
non-equilibrium chemistry, 
and/or stellar irradiation will be presented in future works. 
In particular, current state-of-the-art cloudy boundary-condition models, such as Sonora-diamondback \citep{Morley2024}, primarily focus on silicate clouds relevant to hotter substellar atmospheres. However, for the late-stage evolution of lower-mass ($\lesssim 10,M_J$) gas giants, water and ammonia clouds are expected to play a more dominant role \citep{Chen2023}. 
Incorporating these cloud species will be a key focus of our next release.



\section*{Acknowledgments}

We thank Brianna Lacy for updated opacity tables and helpful discussions. 
Partial funding for this research was provided by the Center for Matter at Atomic Pressures (CMAP), a National Science Foundation (NSF) Physics Frontier Center, under Award PHY-2020249. 
Any opinions, findings, conclusions or recommendations expressed in this material are those of the 
author(s) and do not necessarily reflect those of the National Science Foundation.

%

\vspace{5mm}


\software{\texttt{CoolTLusty} \citep{Hubeny1995,Sudarsky2000,Sudarsky2003,Sudarsky2005,Burrows2008}. The boundary condition data are publicly available at Zenodo at \href{https://zenodo.org/records/18832237?preview=1&token=eyJhbGciOiJIUzUxMiJ9.eyJpZCI6IjUzZDUwMDg5LWEzMmUtNDA1NC1hY2EzLTY0ODM2N2Y2ZjQ2YiIsImRhdGEiOnt9LCJyYW5kb20iOiIwZWM5MmFmMDgxYjg0NTZkYWMyOTJlNTVhNmJhMGVhNSJ9.SJzNTSibv_p7mU40zK5PYApwaUBJLhMPHJYeP0AC0mhl5_3VaGwroZRwn0HwkiMt5JbUtoKmWwrAKEqQpVIVBA}{this link}, where we also append a Python file and detailed instructions to access and interpolate the data.
}





\bibliography{sample631}{}
\bibliographystyle{aasjournal}



\end{document}